\documentclass[pra,twocolumn,showpacs,superscriptaddress,floatfix,nofootinbib]{revtex4}
\usepackage{epsfig}
\usepackage{latexsym}
\usepackage{amsmath}
\usepackage{graphics}
\newcommand{\br}{\mathbf{r}}
\newcommand{\bk}{\mathbf{k}}
\newcommand{\brp}{\mathbf{r}_\parallel}
\newcommand{\bkp}{\mathbf{k}_\parallel}
\newcommand{\TE}{{\rm TE}}
\newcommand{\TM}{{\rm TM}}

\newcommand{\rd}{\hspace{-1mm}{\rm d}}
\newcommand{\w}{\omega}
\newcommand{\curl}{\boldsymbol{\nabla}\times}
\newcommand{\diw}{\boldsymbol{\nabla}\cdot}
\begin{document}

\title{Gauge transformation in macroscopic quantum electrodynamics\\ near polarizable surfaces}
\author{Robert Zietal}
\affiliation{School of Mathematics and Physics, The University of Queensland, Brisbane, Queensland 4072, Australia}
\author{Claudia Eberlein}
\affiliation{School of Science, Loughborough University, Leicestershire, LE11 3TU, England}

\date{\today}
\begin{abstract}
To describe charged particles interacting with the quantized electromagnetic field, we show the differences of working in the so-called generalized and the true Coulomb gauges. We find an explicit gauge transformation between them for the case of the electromagnetic field operators quantized near a macroscopic boundary described by a piece-wise constant dielectric function. Starting from the generalized Coulomb gauge we transform operators into the true Coulomb gauge where the vector potential operator is truly transverse everywhere. We find the generating function of the gauge transformation to carry out the corresponding unitary transformation of the Hamiltonian and show that in the true Coulomb gauge the Hamiltonian of a particle near a polarizable surface contains extra terms due to the fluctuating surface charge density induced by the vacuum field. This extra term is represented by a second-quantised operator on equal footing with the vector field operators. We demonstrate that this term contains part of the electrostatic energy of the charged particle interacting with the surface and that the gauge invariance of the theory guarantees that the total interaction energy in all cases equals the well known result obtainable by the method of images when working in generalized Coulomb gauge. 
The mathematical tools we have developed allow us to work out explicitly the equal-time commutation relations and shed some light on typical misconceptions regarding issues of whether the presence of the boundaries should affect the field commutators or not, especially when the boundaries are modelled as perfect reflectors. 
\end{abstract}

\maketitle

\section{\label{sec:1}Introduction}

Quantum electrodynamics in the presence of polarizable boundaries is a crucial element of the theory describing the interaction of quantum particles with surfaces. It is becoming increasingly relevant thanks to progress in both nanotechnology and experimental techniques in atomic physics. Modern state-of-the-art measurements in atomic physics have reached impressive level of accuracy and subject any theory to previously unparalleled scrutiny \cite{Zimmermann,Harber,Chwedenczuk}. There is a vast number of theoretical tools at one's disposal for studying the interaction between surfaces and quantum objects which can be thought of as being mediated by the electromagnetic field in its vacuum or a thermal state \cite{Bardeen,Casimir,McLachlan,Barton,Power,Sipe,Yeung,Robaschik,Bechler,ActaSlovaca}. All approaches are similar in one aspect, in that they treat the coupling between the quantum object and the quantized electromagnetic field perturbatively. On the other hand, the interaction between the electromagnetic field and the boundary surface needs to be taken into account to all orders, i.e. an exact solution of the operator-valued Maxwell equations is required. It is this aspect of the theory, i.e. the quantization of the electromagnetic field in the presence of a boundary, that even today is a subject of ongoing discussions \cite{Bordag,MiltonNew}. 

Quantum electrodynamics (QED) in free space has been formulated in a variety of ways to suit every need and taste \cite{CT}. Whatever approach is taken, is ultimately dictated merely by convenience, and the gauge invariance of the theory guarantees final results to coincide. However, the situation is different in macroscopic cavity QED where one typically accounts for the presence of the material boundaries by introducing a spatially dependent dielectric function $\epsilon(\br)$, which is usually taken to be a piecewise constant function of position $\br$. Even in the presence of boundaries, the theory still needs to be gauge invariant, but the choice of gauges that are convenient to work with becomes rather restricted \cite{Dalton}. Macroscopic QED is one of the two fundamentally different approaches to formulate the QED in the presence of polarizable media. Another way is to consider the electromagnetic field as interacting with the microscopic constituents of the macroscopic body, which is essentially equivalent, at least at the initial stage when the theory is formulated, to QED in free space. Throughout this article we will focus strictly on the macroscopic approach to cavity QED of non-relativistic particles i.e. those describable by the Schr\"odinger equation. 

For a quantum theory to be consistently formulated one needs a Hamiltonian and an appropriate set of commutation relations between canonically conjugate variables, which in turn allow Heisenberg equations of motion to be derived. In the case of QED a clear-cut way of achieving this is to start from a classical Lagrangian that yields the macroscopic Maxwell equations. At this point, for suitably chosen generalised coordinates, one can unambiguously identify canonically conjugate momenta and proceed to write down the Hamiltonian by using a Legendre transformation. Quantization is then achieved by the correspondence principle, i.e. by converting Poisson brackets to commutators. Formally the transition from the Lagrangian, usually written down in terms of electromagnetic potentials $\mathbf{A}(\br,t)$ and $\phi(\br,t)$, to the Hamiltonian does not require any specific gauge to be chosen. The Hamiltonian may be written in a gauge invariant form, that is, in terms of the electric and magnetic fields alone \cite{Babiker}. However, there is a price to be paid for that, which is that the Hamiltonian takes on a superficial form in which the coupling terms are not manifestly apparent. Thus, for most practical purposes, e.g. in order to apply perturbation theory, a specific gauge needs to be chosen, which in turn affects the workable form of the Hamiltonian. The choice of the gauge is usually restricted to the ones in which an explicit form of the non-interacting electromagnetic operators is easily derived. A common aim is to decouple equations of motion for the potentials $\mathbf{A}(\br,t)$ and $\phi(\br,t)$ and deal with them separately. For QED in free space, this can be achieved in a number of gauges, the most popular being the Coulomb [$\boldsymbol{\nabla}\cdot\mathbf{A}(\br,t)=0$] and the Lorenz [$\boldsymbol{\nabla}\cdot\mathbf{A}(\br,t)+\partial{\phi}(\br,t)/\partial t/c^2=0$] gauge \cite{Jackson}. However, when a polarizable boundary is present and accounted for by introducing piecewise-constant dielectric function $\epsilon=\epsilon(\br)$, neither the Coulomb nor the Lorenz gauge result in the decoupling of the equations of motion for $\mathbf{A}(\br,t)$ and $\phi(\br,t)$. Instead, one is led to introduce the so-called generalized Coulomb gauge $\boldsymbol{\nabla}\cdot[\epsilon(\br)\mathbf{A}(\br,t)]=0$, which allows one to retain certain analogies between free-space QED in the Coulomb gauge and macroscopic QED in the presence of boundaries \cite{Glauber}. In particular, in both cases the scalar potential is not quantized and remains static $\phi(\br,t)=\phi(\br)$ for a static charge distribution. This yields the instantaneous Coulomb interaction between free charges. However, in the case of macroscopic QED with boundaries, this interaction also includes the coupling of charges to the surface, which for simple enough geometries can be determined by the method of images \cite{Jackson}. 

This paper demonstrates how to arrive consistently at a correct formulation of QED in the presence of a polarizable boundary in the \emph{true} Coulomb gauge. This is done by finding an explicit gauge transformation connecting the generalized Coulomb gauge $\boldsymbol{\nabla}\cdot[\epsilon(\br)\mathbf{A}(\br,t)]=0$ with the true Coulomb gauge $\boldsymbol{\nabla}\cdot\mathbf{A}(\br,t)=0$. It will be shown how the Hamiltonian in the true Coulomb gauge can be obtained from that in the generalized Coulomb gauge by a unitary transformation. Once the Hamiltonian is known, one can use standard perturbation theory to calculate interaction energies between charges and surfaces; we shall be demonstrating the gauge invariance of macroscopic QED by explicit calculation of the electrostatic contributions to the interaction of an electron with dielectric surface. The explicit connection we make between the generalized and true Coulomb gauges is very useful in practical calculations because, while  the generalized Coulomb gauge is where the field equations can easily be solved, the true Coulomb gauge is one where a wealth of knowledge exists on how to develop and handle interacting quantum field theories.

\section{\label{sec:GC}Generalised Coulomb gauge}
Although the considerations we report here are quite general, we would like to explain them by referring to a specific example. To that end, we consider a dielectric half-space occupying the region of space $z<0$. For simplicity, the dielectric is assumed to be non-dispersive, i.e. its electromagnetic response is described by a single number, the index of refraction $n$, that is one and the same for all frequencies. This model is described by the dielectric constant
\begin{equation}
\epsilon(z)=1+\theta(-z)(n^2-1)\label{eqn:Epsilon}
\end{equation}
where $\theta(z)$ is the Heaviside step function. The quantization of the electromagnetic field that coexists with such a dielectric can be achieved by normal-mode expansion \cite{Glauber}. We start with Maxwell's equations without sources, 
\begin{eqnarray}
\diw\mathbf{D}(\br,t)&=&0,\label{eqn:Max1}\\
\diw\mathbf{B}(\br,t)&=&0,\label{eqn:Max2}\\
\curl\mathbf{E}(\br,t)+\frac{\partial}{\partial t}\mathbf{B}(\br,t)&=&0,\label{eqn:Max3}\\
\curl\mathbf{H}(\br,t)-\frac{\partial}{\partial t}\mathbf{D}(\br,t)&=&0.\label{eqn:Max4}
\end{eqnarray} 
For a material that is non-magnetic and has the non-dispersive dielectric function (\ref{eqn:Epsilon}), the constitutive relations may be written as
\begin{equation}
\mathbf{B}(\br,t)=\mu_0\mathbf{H}(\br,t),\;\;\mathbf{D}(\br,t)=\epsilon_0\epsilon(z)\mathbf{E}(\br,t) 
\end{equation}
Introducing the electromagnetic potentials in the usual way \cite{Jackson}
\begin{eqnarray}
\mathbf{E}(\br,t)=-\frac{\partial}{\partial t}\mathbf{A}(\br,t)-\boldsymbol{\nabla}\phi(\br,t)\label{eqn:potE}\\
\mathbf{B}(\br,t)=\curl\mathbf{A}(\br,t)\label{eqn:potB},
\end{eqnarray}
takes care of Eqs.~(\ref{eqn:Max2}) and (\ref{eqn:Max3}). The remaining two Maxwell equations (\ref{eqn:Max1}) and (\ref{eqn:Max4}) turn into:
\begin{eqnarray}
\diw\left[\epsilon(z)\boldsymbol{\nabla}\phi(\br,t)\right]+\frac{\partial}{\partial t}\diw\left[\epsilon(z)\mathbf{A}(\br,t)\right]&=&0,\label{eqn:Wave1}\\
\curl[\curl\mathbf{A}(\br,t)]+\frac{\epsilon(z)}{c^2}\frac{\partial^2}{\partial t^2}\mathbf{A}(\br,t)\nonumber\\
+\frac{\epsilon(z)}{c^2}\frac{\partial}{\partial t}\boldsymbol{\nabla}\phi(\br,t)&=&0.\label{eqn:Wave2}
\end{eqnarray}
The solution of these coupled differential equations can be very much simplified by a suitable choice of gauge for the electromagnetic potentials. It is expedient to decouple the two equations. In non-relativistic QED, the most convenient approach is to work in the generalized Coulomb gauge where we require that
\begin{eqnarray}
\diw\left[\epsilon(z)\mathbf{A}(\br,t)\right]&=& 0 ,\nonumber\\
\epsilon(z)\diw\mathbf{A}(\br,t)+(1-n^2)A_z(\br,t)\delta(z)&=&0.\;\label{eqn:Generalized}
\end{eqnarray}
where the specific form of the dielectic constant of Eq.~(\ref{eqn:Epsilon}) has been used to get the second line. We note that, since $\epsilon(z)$ is not spatially uniform but has a finite jump at $z=0$, the generalised Coulomb gauge differs from the standard Coulomb gauge
\begin{equation}
\diw\mathbf{A}(\br,t)=0\label{eqn:TrueCoulomb}
\end{equation}
by a surface term that is proportional to a $\delta(z)$-function. With Eq.~(\ref{eqn:Generalized}) it follows from Eq.~(\ref{eqn:Wave1}) that in the absence of sources we can set $\phi(\br,t)=0$. Thus in generalized Coulomb gauge, Eq.~(\ref{eqn:Wave2}) reduces to
\begin{equation}
\curl[\curl\mathbf{A}(\br,t)]+\frac{\epsilon(z)}{c^2}\dfrac{\partial^2}{\partial t^2}\mathbf{A}(\br,\w)=0.
\end{equation}
Therefore, only the vector potential undergoes quantization, which is accomplished by expanding $\mathbf{A}(\br,t)$ in a complete set of the mode functions that satisfy
\begin{equation}
\curl[\curl\mathbf{f}_\sigma(\br)]-\epsilon(z)\dfrac{\w_\sigma^2}{c^2}\mathbf{f}_\sigma(\br)=0,
\end{equation}
and are supplemented by the condition that derives from the gauge we are working in, cf. Eq.~(\ref{eqn:Generalized})
\begin{equation}
\diw\left[\epsilon(z)\mathbf{f}_\sigma(\br)\right]=0.\label{eqn:GaugeForModes}
\end{equation}
We have labelled solutions corresponding to the eigenvalue $\w_\sigma$ by $\sigma$. The double-curl operator can be rewritten using Eq.~(\ref{eqn:GaugeForModes})
\begin{eqnarray}
\curl[\curl\mathbf{f}_\sigma(\br)]&=&\boldsymbol{\nabla}\left[\diw\mathbf{f}_\sigma(\br)\right]-\nabla^2\mathbf{f}_\sigma(\br,\sigma)\nonumber\\
&=&-\nabla^2\mathbf{f}_\sigma(\br,\sigma),\hspace{5mm}\mbox{for}\; z\neq0.\;\;\;\nonumber
\end{eqnarray}
Thus away from the interface we can work with the Helmholtz equation 
\begin{equation}
\nabla^2\mathbf{f}_\sigma(\br)+\epsilon(z)\dfrac{\w_\sigma^2}{c^2}\mathbf{f}_\sigma(\br)=0,\;\;\;\mbox{for}\; z\neq 0,
\end{equation} 
which can be solved as usual by considering the two distinct regions of space, $z<0$ and $z>0$, and using Maxwell boundary conditions to match solutions across the interface. Once the mode functions are known, the expansion of the vector potential is written as
\begin{equation}
\mathbf{A}^{\rm gc}(\br,t)=\sum_\sigma\sqrt{\frac{\hbar}{2\epsilon_0\w_{\sigma}}}\left[a_\sigma\mathbf{f}_\sigma(\br)e^{-i\w_\sigma t}+{\rm C.C.}\right],\label{eqn:AExpansion}
\end{equation}
where the superscript gc reminds us that the expansion is written down in generalized Coulomb gauge, Eq.~(\ref{eqn:Generalized}). Quantization is accomplished by the promotion of the expansion coefficients $a_\sigma$ to operators that satisfy bosonic equal-time commutation rules
\begin{eqnarray}
&&[\hat{a}_\sigma, \hat{a}^\dagger_{\sigma'}]=\delta_{\sigma, \sigma'},\\
&&[\hat{a}_\sigma, \hat{a}_{\sigma'}]=0. \nonumber
\end{eqnarray}
In the present geometry, described by the dielectric function (\ref{eqn:Epsilon}), the procedure outlined above yields the well-known Carnigila-Mandel modes for the vector field operator which naturally split into two parts describing left-incident and right-incident photons, respectively \cite{Carnigila}:
\begin{widetext}
\begin{eqnarray}
\hat{\mathbf{A}}^{\rm gc}(\br,t)&=&\sum_\lambda\int\rd^2\bkp
\left\{\left[\int_0^\infty \rd k_{zd}\sqrt{\frac{\hbar}{2\epsilon_0\w_{\bk\lambda}}}\;\hat{a}^{L}_{\bk\lambda}(t)\mathbf{f}_{\bk\lambda}^L(\br)\right]+\left[\int_0^\infty\rd k_{z}\sqrt{\frac{\hbar}{2\epsilon_0\w_{\bk\lambda}}}\;\hat{a}^{R}_{\bk\lambda}(t)\mathbf{f}_{\bk\lambda}^R(\br)\right]\right\}+{\rm H.C.}\;\;\;\label{eqn:AOperator}\\
\mathbf{f}^L_{\bk\lambda}(\br)&=&\dfrac{\hat{\mathbf{e}}_\lambda(\boldsymbol{\nabla})}{(2\pi)^{3/2}n}\left\{ \theta(-z)\left[e^{i\bk^+_d\cdot\br}+R_\lambda^L e^{i\bk^-_d\cdot\br}\right]+ \theta(z)\left[T^L_\lambda e^{i\bk^+\cdot\br} \right]\right\}\label{eqn:LeftIncident}\\
\mathbf{f}^R_{\bk\lambda}(\br)&=&\dfrac{\hat{\mathbf{e}}_\lambda(\boldsymbol{\nabla})}{(2\pi)^{3/2}}\left\{ \theta(z)\left[e^{i\bk^-\cdot\br}+R_\lambda^R e^{i\bk^+\cdot\br}\right]+ \theta(-z)\left[T^R_\lambda e^{i\bk_d^+\cdot\br} \right]\right\}\label{eqn:RightIncident}
\end{eqnarray}
\end{widetext}
Here $\lambda$ labels the polarization of the photons ${\lambda=\{\TE, \TM \}}$ as transverse electric and transverse magnetic, and a harmonic time-dependence of the annihilation and creation operators is implicitly assumed i.e. ${a_{\bk\lambda}(t)=a_{\bk\lambda}(0)e^{-i\w_{\bk\lambda}t}}$. The mode functions $\mathbf{f}_{\bk\lambda}(\br)$ entering the expansion (\ref{eqn:AOperator}) contain wavevectors $\bk$ and $\bk_d$ i.e. the wavevectors in vacuum and dielectric, respectively
\begin{equation}
\bk^\pm=(\bkp,\pm k_z),\;\;\bk_d^\pm=(\bkp,\pm k_{zd}).
\end{equation}
Their $z$-components are related to each other via the law of refraction, ${k_{zd}=\sqrt{n^2k_z^2+(n^2-1)\bkp^2}}$. The sign of the square root is chosen in such a way that on the real axis we have ${\rm sgn}(k_{z})={\rm sgn}(k_{zd})$. This ensures that for a single mode of the electromagnetic field that consists of incident, reflected and transmitted waves, the direction of propagation is consistent between those waves. In Eqs.~(\ref{eqn:LeftIncident}) and (\ref{eqn:RightIncident}) a shorthand notation has been introduced to represent the unit polarization vectors $\hat{\mathbf{e}}_\lambda$. We have defined them as
\begin{eqnarray}
\hat{\mathbf{e}}_\TE(\boldsymbol{\nabla})&=&
\left(-\nabla^2_\parallel\right)^{-1/2}\left(-i\nabla_y,i\nabla_x,0\right), \label{eqn:PW1}\\
\hat{\mathbf{e}}_\TM(\boldsymbol{\nabla})&=&
\left(\nabla^2_\parallel\nabla^2\right)^{-1/2}\left(-\nabla_x\nabla_z,-\nabla_y\nabla_z,\nabla^2_\parallel\right), \label{eqn:PW2}
\end{eqnarray}
where it is understood that the derivatives are acting on plane waves and thus give the corresponding components of the wave vector of that wave, e.g. for the right-incident incoming wave $e^{i\bk^-\cdot\br}$ the operator $\nabla_z$ gives $-ik_z$. We emphasize that our notation is such that the polarization vectors do not act on the step function in $\epsilon(z)$. This is a convenient notation as the polarization vectors point in different directions for incident, reflected and transmitted waves, respectively. However, one needs to be careful when carrying out explicit calculations with the mode functions (\ref{eqn:LeftIncident})--(\ref{eqn:RightIncident}) and remember that the operator $\hat{\mathbf{e}}_\lambda(\boldsymbol{\nabla})$ is merely a shorthand notation. The Fresnel coefficients in mode functions (\ref{eqn:LeftIncident}) and (\ref{eqn:RightIncident}) are given by
\begin{eqnarray}
R^R_{\TE}=\frac{k_z-k_{zd}}{k_z+k_{zd}},\;\;R^R_{\TM}=\frac{n^2k_z-k_{zd}}{n^2k_z+k_{zd}},\;\;R^L_{\lambda}=-R^R_{\lambda},\nonumber\\
T^R_{\TE}=\frac{2k_{z}}{k_z+k_{zd}},\;\;T^R_{\TM}=\frac{2nk_{z}}{n^2k_z+k_{zd}},\;\;T^L_\lambda=\dfrac{k_{zd}}{k_{z}}T^R_\lambda.\nonumber\\\label{eqn:Frnl}
\end{eqnarray}
The mode functions (\ref{eqn:LeftIncident})--(\ref{eqn:RightIncident}) need to satisfy a completeness relation which can be written in the form \cite{Glauber} 
\begin{eqnarray}
\sum_\lambda\int\rd^2\bkp\bigg[\int_0^\infty \rd k_z \;f_{\bk\lambda, i}^R(\br)f_{\bk\lambda, j}^{*R}(\br')\hspace{25mm}\nonumber\\
+\int_0^\infty \rd k_{zd} \;f_{\bk\lambda, i}^L(\br)f_{\bk\lambda, j}^{*L}(\br')
\bigg]
=\delta^\epsilon_{ij}(\br,\br')\hspace{0.2 cm}\label{eqn:Completeness}
\end{eqnarray}
where for definiteness throughout this paper we choose $\br'$ to refer to a point that lies outside dielectric, i.e. $z'>0$. 
The proof of the relation
\begin{eqnarray}
\nabla^2\sum_\lambda\int\rd^2\bkp\bigg[\int_0^\infty \rd k_z \;f_{\bk\lambda, i}^R(\br)f_{\bk\lambda, j}^{*R}(\br')\hspace{1.5cm}\nonumber\\
+\int_0^\infty \rd k_{zd} \;f_{\bk\lambda, i}^L(\br)f_{\bk\lambda, j}^{*L}(\br')
\bigg]\hspace{2.0 cm}\nonumber\\
=\left(\nabla_i\nabla_j-\delta_{ij}\nabla^2\right)\delta^{(3)}(\br-\br')\label{eqn:Completeness1}
\end{eqnarray}
has been presented in \cite{Birula}. Equation (\ref{eqn:Completeness1}) is of course obtained by acting with the Laplace operator $\nabla^2$ on (\ref{eqn:Completeness}). However, at this point it is not obvious that
\begin{equation}
\nabla^2\delta_{ij}^{\epsilon}(\br,\br')=\left(\nabla_i\nabla_j-\delta_{ij}\nabla^2\right)\delta^{(3)}(\br-\br').\label{eqn:BirulaDelta}
\end{equation}
The object $\delta_{ij}^{\epsilon}(\br,\br')$ represents the unit kernel in the subspace of the mode functions that satisfy the generalised Coulomb gauge i.e. if $\mathbf{f}_{\bk\lambda}(\br)$ satisfies Eq. (\ref{eqn:GaugeForModes}) then
\begin{equation}
\int\rd ^3\br'\delta^\epsilon_{ij}(\br,\br')\mathbf{f}^j_{\bk\lambda}(\br')=\mathbf{f}^i_{\bk\lambda}(\br).
\end{equation}
Even less obvious is that, even though the generalized Coulomb gauge differs from the standard Coulomb gauge only by a surface term, cf. Eq. (\ref{eqn:Generalized}), the corresponding unit kernels in the position representation in these two gauges differ in the whole of space because of their non-local character, i.e. even though
\begin{equation}
\diw \mathbf{f}_{\bk\lambda}(\br)=\diw \left[\epsilon(z)\mathbf{f}_{\bk\lambda}(\br)\right],\;{\rm for}\;z\neq0,
\end{equation}
we have 
\begin{equation}
\delta^\perp_{ij}(\br-\br')\neq\delta^\epsilon_{ij}(\br,\br'),\;\;\;{\rm for\;all}\;z,z'.
\end{equation}
Here, $\delta^{\perp}_{ij}(\br-\br')$ is the usual transverse $\delta$-function
\begin{equation}
\delta^{\perp}_{ij}(\br-\br')=\frac{1}{(2\pi)^3}\int\rd^3\bk\left(\delta_{ij}-\frac{k_i k_j}{\bk^2}\right)e^{i\bk\cdot(\br-\br')},\label{eqn:TransverseDelta}
\end{equation}
i.e. the unit kernel in the subspace of mode functions that satisfy $\diw \mathbf{f}_{\bk\lambda}(\br)=0$. We also emphasize that $\delta^\epsilon_{ij}(\br,\br')$ is not translation-invariant, because translation invariance is broken by the presence of the interface where waves are partially reflected.

\begin{figure}[htbp]
  \centering
    \includegraphics[width=8cm,height=6cm]{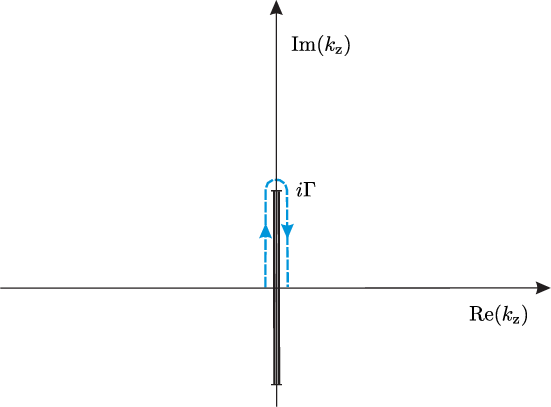}
  \caption{The dashed line represents the contour $\mathcal{C}$ used to evaluate the $k_z$-integral in equation (\ref{eqn:AroundCut1}).}
  \label{fig:AroundCut}
\end{figure}

It turns out that it is possible to calculate the $\br$-representation of $\delta^\epsilon_{ij}(\br,\br')$ directly by evaluating the integrals in (\ref{eqn:Completeness}). Before we do so, let us rewrite the transverse delta function (\ref{eqn:TransverseDelta}) as
\begin{equation}
\delta^{\perp}_{ij}(\br-\br')=\delta_{ij}\delta^{(3)}(\br-\br')-\nabla_i\nabla_j' G^0(\br-\br'),\label{eqn:TransverseDelta1}
\end{equation}
where we have introduced the Green's function of the Poisson equation in free space
\begin{equation}
G^0(\br-\br')=\frac{1}{4\pi}\frac{1}{|\br-\br'|}\;.\label{eqn:Poisson}
\end{equation} 
Let us now turn to the explicit evaluation of the LHS of Eq. (\ref{eqn:Completeness}). First we deal with the case $z<0$ and $z'>0$ for which we provide a detailed calculation. Substituting the mode functions (\ref{eqn:LeftIncident})--(\ref{eqn:RightIncident}) into (\ref{eqn:Completeness}) and multiplying out we obtain
\begin{eqnarray}
\delta_{ij}^\epsilon(\br,\br')=\frac{1}{(2\pi)^3}\sum_\lambda\int \rd^2\bkp e^{i\bkp\cdot(\brp-\brp')}\hspace{1cm}
\nonumber\\
\times
\left\{\int_{0}^\infty \dfrac{\rd k_{zd}}{n^2}\left[ T^{L*}_\lambda\;\hat{e}^i_\lambda(\bk_d^+)\hat{e}^{*j}_{\lambda}(\bk^+)e^{ik_{zd} z-ik_z^*z'}
\right.\right.
\nonumber\\
\left.\left.
+R^{L}_\lambda T^{L*}_\lambda\; \hat{e}^i_\lambda(\bk_d^-)\hat{e}^{*j}_{\lambda}(\bk^+)e^{-ik_{zd}z-ik_z^*z'}\right]\right.
\nonumber\\
\left.+\int_0^\infty \rd k_z \left[ T^{R}_\lambda \;\hat{e}^i_\lambda(\bk_d^-)\hat{e}^{*j}_{\lambda}(\bk^-)e^{-ik_{zd}z+ik_zz'}
\right.\right.
\nonumber\\
\left.\left.
+R^{R}_\lambda T^{R}_\lambda \;\hat{e}^i_\lambda(\bk_d^-)\hat{e}^{*j}_{\lambda}(\bk^+)e^{-ik_{zd}z-ik_zz'}\right]\right\}\nonumber\\
\label{eqn:CompExplicitA1}
\end{eqnarray}
where $\hat{e}_\lambda^i(\bk^\pm)\equiv \hat{e}_\lambda^i(\boldsymbol{\nabla}) e^{i\bk^\pm\cdot\br}$. We proceed by focussing attention on the $k_z$ and $k_{zd}$ integrals. We convert the $k_{zd}$ integral using the relation ${k_{zd}=\sqrt{n^2k_z^2+(n^2-1)\bkp^2}}$
\begin{equation}
\int_0^\infty\rd k_{zd} = n^2\int_{i\Gamma}^0\rd k_z \dfrac{k_z}{k_{zd}}+n^2\int_0^\infty\rd k_z\dfrac{k_z}{k_{zd}},
\end{equation}
where  $\Gamma=|\bkp|(n^2-1)^{1/2}/n$. After this change of variables the expression we wish to evaluate consists of an integral along the real-positive axis (travelling modes) and an integral along part of the positive imaginary axis where $k_z \in [0,\Gamma]$ (evanescent modes)
\begin{eqnarray}
\delta_{ij}^\epsilon(\br,\br')=\frac{1}{(2\pi)^3}\sum_\lambda\int \rd^2\bkp e^{i\bkp\cdot(\brp-\brp')}\hspace{1.0 cm}\nonumber\\
\times
\left\{
\int_{i\Gamma}^{0^+} \rd k_z
\left[ 
\dfrac{k_z}{k_{zd}}T^{L*}_\lambda\;\hat{e}^i_\lambda(\bk_d^+)\hat{e}^{j}_{\lambda}(\bk^-)e^{ik_{zd} z+ik_z z'}
\right.\right.
\nonumber\\
\left.\left.
+ T^{L*}_\lambda R^{L}_\lambda\; \hat{e}^i_\lambda(\bk_d^-)\hat{e}^{j}_{\lambda}(\bk^-)e^{-ik_{zd}z+ik_zz'}
\right]
\right.\nonumber\\
\left.
+\int_{0}^\infty \rd k_z
\left[ 
\dfrac{k_z}{k_{zd}}T^{L}_\lambda\;\hat{e}^i_\lambda(\bk_d^+)\hat{e}^{j}_{\lambda}(\bk^+)e^{ik_{zd} z-ik_z z'}
\right.\right.
\nonumber\\
\left.\left.
+ T^{R}_\lambda\; \hat{e}^i_\lambda(\bk_d^-)\hat{e}^{j}_{\lambda}(\bk^-)e^{-ik_{zd}z+ik_zz'}
\right.
\right.\nonumber\\
\left. \left.
+\dfrac{k_z}{k_{zd}} T^{L}_\lambda R^L_\lambda  \;\hat{e}^i_\lambda(\bk_d^-)\hat{e}^{j}_{\lambda}(\bk^+)e^{-ik_{zd}z-ik_zz'}
\right.\right.
\nonumber\\
\left.\left.
+R^{R}_\lambda T^{R}_\lambda \;\hat{e}^i_\lambda(\bk_d^-)\hat{e}^{j}_{\lambda}(\bk^+)e^{-ik_{zd}z-ik_zz'}
\right]\right\}.\label{eqn:CompExplicitA2}
\end{eqnarray}
Here the integral on the interval $k_z\in [i\Gamma, 0^+]$ runs on the right side of the branch cut due to $k_{zd}$ that runs from $k_z=-i\Gamma$ to $k_z=i\Gamma$. The last two terms in Eq.~(\ref{eqn:CompExplicitA2}) cancel out by virtue of the relations (\ref{eqn:Frnl}), and the other two terms in that integral can be combined to a single integral running along the interval $k_z\in (-\infty, 0^-] \cap [0^+, \infty)$
\begin{eqnarray}
\delta_{ij}^\epsilon(\br,\br')=\frac{1}{(2\pi)^3}\sum_\lambda\int \rd^2\bkp e^{i\bkp\cdot(\brp-\brp')}\hspace{23mm}\nonumber\\
\times\left\{
\int_{-\infty}^\infty \rd k_z
\left[ 
 T^{R}_\lambda\; \hat{e}^i_\lambda(\bk_d^-)\hat{e}^{j}_{\lambda}(\bk^-)e^{-ik_{zd}z+ik_zz'}
\right]\right.\hspace{2mm}\nonumber\\
\left.
+\int_{i\Gamma}^{0^+} \rd k_z
\left[ 
\dfrac{k_z}{k_{zd}}T^{L*}_\lambda\;\hat{e}^i_\lambda(\bk_d^+)\hat{e}^{j}_{\lambda}(\bk^-)e^{ik_{zd} z+ik_z z'}
\right.\right.\hspace{-2mm}
\nonumber\\
\left.\left.
+ T^{L*}_\lambda R^{L}_\lambda\; \hat{e}^i_\lambda(\bk_d^-)\hat{e}^{j}_{\lambda}(\bk^-)e^{-ik_{zd}z+ik_zz'}
\right]
\right\}.\hspace{4mm}
\label{eqn:CompExplicitA3}
\end{eqnarray}
To proceed any further, close inspection of Eq.~(\ref{eqn:CompExplicitA3}) is necessary. To illustrate the argument, we focus on the TM contributions to the integral. The TE contributions are treated in an exactly analogous way. We start by noting that for purely imaginary $k_z$ we have $k_z^*=-k_z$ so that we get
\begin{equation}
T^{L*}_\TM=\dfrac{2nk_{z}}{k_{zd}-n^2k_z},\hspace{1 cm}\dfrac{k_z}{k_{zd}}T^{L*}_\TM R^L_{\TM}=\dfrac{2nk_{z}}{k_{zd}+n^2k_z}.\nonumber
\end{equation} 
Therefore, the $k_z$-integral in the last two lines of Eq.~(\ref{eqn:CompExplicitA3}) can be written as
\begin{eqnarray}
\int_{i\Gamma}^{0^+} \rd k_z
\left( 
\dfrac{2nk_z}{k_{zd}+n^2k_z}\right)\hat{e}^i_\TM(\bk_d^-)\hat{e}^{j}_{\TM}(\bk^-)e^{-ik_{zd} z+ik_z z'}\nonumber\\
+\int_{i\Gamma}^{0^+} \rd k_z\left(\dfrac{2nk_z}{k_{zd}-n^2k_z}\right) \hat{e}^i_\TM(\bk_d^+)\hat{e}^{j}_{\TM}(\bk^-)e^{+ik_{zd}z+ik_zz'}.\nonumber
\end{eqnarray}
Now we observe that the second integral differs from the first integral only by the sign of $k_{zd}$. This allows us to combine these two integrals into a single contour integral around the branch-cut due to $k_{zd}$  
\begin{eqnarray}
\int_\mathcal{C} \rd k_z
T^R_\TM\hat{e}^i_\TM(\bk_d^-)\hat{e}^{j}_{\TM}(\bk^-)e^{-ik_{zd} z+ik_z z'}\label{eqn:AroundCut1}
\end{eqnarray}
where the contour $\mathcal{C}$ is illustrated in Fig. \ref{fig:AroundCut}. 
Thus the completeness relation (\ref{eqn:CompExplicitA3}) may be written compactly as
\begin{eqnarray}
\delta_{ij}^\epsilon(\br,\br')=\frac{1}{(2\pi)^3}\sum_\lambda\int \rd^2\bkp e^{i\bkp\cdot(\brp-\brp')}\hspace{0.5cm}\nonumber\\
\times\int_\gamma \rd k_z 
 T^{R}_\lambda\; \hat{e}^i_\lambda(\bk_d^-)\hat{e}^{j}_{\lambda}(\bk^-)e^{-ik_{zd}z+ik_zz'}
\label{eqn:CompExplicitA4}
\end{eqnarray}
where the contour $\gamma$ runs along the negative real axis from $k_z=-\infty$ to $k_z=0^-$, then around the branch-cut along the contour $\mathcal{C}$ depicted in Fig. \ref{fig:AroundCut} and then from $k_z=0^+$ to $k_z=\infty$. The $k_z$-integral may now be evaluated with the help of the residue theorem. We note that for $z<0$ and $z'>0$ the integrand in Eq.~(\ref{eqn:CompExplicitA4}) vanishes exponentially in the upper $k_z$-plane so that we can close the contour there. To do so we need to determine the position of the integrand's poles, if any. The Fresnel coefficients for the half-space geometry are analytic for ${\rm Im}(k_z)>0$ so that it remains to look at the analytic properties of the polarization vectors defined in Eqs.~(\ref{eqn:PW1})--(\ref{eqn:PW2}). For the TE mode we immediately note that $\hat{\mathbf{e}}_\TE$ are independent of $k_z$. Thus the transverse electric modes do not contribute to the integral (\ref{eqn:CompExplicitA4}). For the TM mode, each polarization vector contributes a factor of $1/|\bk|$ where $|\bk|=\sqrt{k_z^2+\bkp^2}$. Thus for a TM mode the integrand has a simple pole in the upper half-plane at $k_z=i|\bkp|$. Using the residue theorem, one can now easily show that
\begin{equation}
\delta_{ij}^\epsilon(\br,\br')=-\nabla_i \nabla'_j G^T(\br-\br'),\mbox{ for }z<0,\;z'>0\label{eqn:GeneralizedDeltaLeft}
\end{equation}
where 
\begin{equation}
G^T(\br-\br')=\frac{1}{4\pi n^2}\frac{2n^2}{n^2+1}\frac{1}{|\br-\br'|}
\end{equation}
is the transmitted part of the electrostatic Green's function in the half-space geometry, see e.g. \cite{Jackson}.

In order to evaluate Eq.~(\ref{eqn:Completeness}) for the case $z>0,\;z'>0$ we again substitute the relevant the mode functions (\ref{eqn:LeftIncident})--(\ref{eqn:RightIncident}) and after utilizing straightforward properties of the Fresnel reflection coefficients we arrive at
\begin{eqnarray}
\delta_{ij}^\epsilon(\br,\br')=\frac{1}{(2\pi)^3}\sum_\lambda\int \rd^2\bkp e^{i\bkp\cdot(\brp-\brp')}\hspace{24mm}\nonumber\\
\times\left\{\int_{-\infty}^\infty\rd k_z \hat{e}^i_\lambda(\bk^+)\hat{e}^j_\lambda(\bk^+)e^{ik_z(z-z')}\right.\hspace{12mm}\nonumber\\
+\int_{-\infty}^\infty\rd k_z R_\lambda^R\hat{e}^i_\lambda(\bk^+)\hat{e}^j_\lambda(\bk^-)e^{ik_z(z+z')}
\hspace{4mm}\nonumber\\
\left.+\int_{i\Gamma}^0\rd k_z \frac{k_z}{k_{zd}}\left| T^L_\lambda\right|^2\hat{e}^i_\lambda(\bk^-)\hat{e}^j_\lambda(\bk^-)e^{ik_z(z+z')}\right\}\; \hspace{2mm}
\label{eqn:CompExplicit}
\end{eqnarray}
with $\Gamma=|\bkp|(n^2-1)^{1/2}/n$ and $\hat{e}_\lambda^i(\bk^\pm)\equiv \hat{e}_\lambda^i(\boldsymbol{\nabla}) e^{i\bk^\pm\cdot\br}$. Now we note that, because of the completeness properties of the polarization vectors, the first $k_z$ integral in Eq.~(\ref{eqn:CompExplicit}) yields the transverse $\delta$-function, Eq. (\ref{eqn:TransverseDelta}). The remaining two terms can be combined into a single contour integral around the branch cut due to $k_{zd}=\sqrt{n^2k_z^2+(n^2-1)\bkp^2}$. This is done in exactly the same manner as in \cite{Birula,Robaschik}. Thus the result reads
\begin{eqnarray}
\delta_{ij}^\epsilon(\br,\br')=\delta^\perp_{ij}(\br-\br')+\frac{1}{(2\pi)^3}\sum_\lambda\int \rd^2\bkp e^{i\bkp\cdot(\brp-\brp')}\nonumber\\
\times\int_\gamma\rd k_z R_\lambda^R\hat{e}^i_\lambda(\bk^+)\hat{e}^j_\lambda(\bk^-)e^{ik_z(z+z')}\;\;
\label{eqn:CompExplicit1}
\end{eqnarray}
where the contour $\gamma$ runs along the negative real axis from $k_z=-\infty$ to $k_z=0^-$, then around the branch cut along the contour $\mathcal{C}$ depicted in Fig. \ref{fig:AroundCut} and then from $k_z=0^+$ to $k_z=\infty$. Since the reflection coefficient $R^R_\lambda$ has no poles in the upper $k_z$-plane we can close the contour there. Then, for the $\TE$ mode the integral vanishes because the polarization vectors do not depend on $k_z$. For the $\TM$ mode, however, the polarization vectors contribute a pole in the upper half-plane at $k_z=i|\bkp|$. The integral is easily evaluated using the residue theorem and leads to the final result that can be written explicitly as
\begin{eqnarray}
\delta_{ij}^\epsilon(\br,\br')&\!\!=&\!\!\delta_{ij}\delta^{(3)}(\br-\br')\nonumber\\
&&\!\!-\nabla_i\nabla'_j\left[G^0(\br-\br')+G^R(\br,\br')\right]\mbox{ for }z,z'>0\nonumber\\\label{eqn:GeneralizedDeltaRight}
\end{eqnarray}
with $G^{\rm R}(\br,\br')$ being the reflected part of the electrostatic Green's function in the half-space geometry 
\begin{eqnarray}
G^{R}(\br,\br')=-\frac{1}{4\pi}\frac{n^2-1}{n^2+1}\frac{1}{|\br-\bar{\br}'|}\label{eqn:ReflectedPoisson}
\end{eqnarray}
where $\bar{\br}'=(x',y',-z')$. 

The results (\ref{eqn:GeneralizedDeltaLeft}) and (\ref{eqn:GeneralizedDeltaRight}) may be written in compact form as
\begin{equation}
\delta_{ij}^\epsilon(\br,\br')=\delta_{ij}\delta^{(3)}(\br-\br')
-\nabla_i\nabla'_j G(\br,\br'),\mbox{ for }z'>0\label{eqn:GeneralizedDeltaTot}
\end{equation} 
where
\begin{eqnarray}
G(\br,\br')&=&\frac{1}{4\pi n^2}\frac{2n^2}{n^2+1}\frac{1}{|\br-\br'|}\theta(-z)\nonumber\\
&+&\left(\frac{1}{4\pi}\frac{1}{|\br-\br'|}-\frac{1}{4\pi}\frac{n^2-1}{n^2+1}\frac{1}{|\br-\bar{\br}'|}\right)\theta(z)\nonumber\\
\label{eqn:FullPoissonGreenF}
\end{eqnarray}
is the Green's function of the Poisson equation for the case of a source being outside the dielectric occupying the $z<0$ region of space. We see that the end result has formally the same form as (\ref{eqn:TransverseDelta1}) only that the free-space Green's function of the Poisson equation is replaced by the Green's function in the presence of a dielectric half-space of refractive index $n$. The result (\ref{eqn:GeneralizedDeltaTot}) may be formally written as
\begin{equation}
\delta_{ij}^\epsilon(\br,\br')=\left(\delta_{ij}+\nabla_i\nabla'_j\nabla^{-2}\right)\delta^{(3)}(\br-\br')\label{eqn:GeneralizedDelta}
\end{equation}
provided an appropriate meaning is attached to the integral operator $\nabla^{-2}$. We would like to remark that it is in this sense that the completeness relation proven in \cite{Completeness} holds. There, of course, the Green's function is that in the slab geometry, see the appendix of Ref.~\cite{Slab}. Equation (\ref{eqn:GeneralizedDelta}) needs to be compared with Eq.~(\ref{eqn:BirulaDelta}). Note in particular, that the derivative $\nabla'_j$ which acts on $\br'$ can not be shifted to act on $\br$ because of the reflection term in (\ref{eqn:FullPoissonGreenF}). This is possible only after one acts with Laplace operator on (\ref{eqn:GeneralizedDelta}). Only then one can replace $\nabla_j'$ with $-\nabla_j$ and recover the result (\ref{eqn:BirulaDelta}) derived in \cite{Birula}.

Once the completeness relation of the mode functions has been explicitly calculated, one can also evaluate the equal-time field commutator. Using Eq.~(\ref{eqn:AExpansion}) we have
\begin{equation}
\left[A_i^{\rm gc}(\br),\epsilon_0 E_j(\br')\right]=-i\hbar\delta^{\epsilon}_{ij}(\br,\br')
\end{equation}
so for the case of the electromagnetic field in the presence of a dielectric half-space the commutator between the vector potential and electric field operator reads
\begin{eqnarray}
\left[A_i^{\rm gc}(\br),\epsilon_0 E_j(\br')\right]&=&-i\hbar\delta_{ij}\delta^{(3)}(\br-\br')\nonumber\\
&+&i\hbar\nabla_i\nabla'_j G(\br,\br').\label{eqn:CommGenGauge}
\end{eqnarray}
where $G(\br,\br')$ is given by (\ref{eqn:FullPoissonGreenF}) and we remind the reader that we consider the case $z'>0$ only. We see that, compared to the standard commutation relations of QED, the commutator in the presence of the dielectric gains an additional term that represents the reflection from the surface. Note that in the limit of perfect reflectivity, i.e. $n\rightarrow\infty$, we recover the results obtained in \cite{Milonni3,Power}. We will come back to this fact at the end of the section \ref{sec:3}.

\section{\label{sec:3}Coulomb gauge}
The natural question arising is whether it is possible to quantize the electromagnetic field in the presence of a dielectric half-space but work in true Coulomb gauge. The direct approach to solving the Maxwell equations (\ref{eqn:Wave1})--(\ref{eqn:Wave2}) proves intractable, but we shall show that one can exploit a gauge transformation for working out the field operators in the true Coulomb gauge from the ones in the generalized Coulomb gauge. A gauge transformation from the generalized Coulomb gauge to the true Coulomb gauge may be written as follows
\begin{eqnarray}
\mathbf{A}^{\rm c}(\br,t)&=&\mathbf{A}^{\rm gc}(\br,t)-\boldsymbol{\nabla}\chi(\br,t),\label{eqn:ChangeA}\\
\phi^{\rm c}(\br,t)&=&\phi^{\rm gc}(\br,t)+\frac{\partial}{\partial t}\chi(\br,t)\label{eqn:changePhi}.
\end{eqnarray}
where we set $\phi^{\rm gc}(\br,t)=0$ in the absence of charges. It is clear that in the true Coulomb gauge, even in the absence of charges, the scalar potential does not vanish. In fact, we shall see shortly that in true Coulomb gauge the scalar potential enters the Hamiltonian on an equal footing with the vector potential as a second-quantized operator. We note that the left-hand side of Eq.~(\ref{eqn:ChangeA}) is transverse, and since $\mathbf{A}^{\rm gc}$ is not, the gradient of the generating function $\chi(\br,t)$ must compensate for it \cite{Babiker}. In other words we have \cite{maximumPrincipleFootnote}
\begin{equation}
\nabla_i \chi(\br,t)=\int\rd^3 \br' \delta_{ij}^\parallel(\br-\br')A_j^{\rm gc}(\br', t).\label{eqn:ChiConstr}
\end{equation}
The form of the $\chi$ can be easily found if we use the position representation of the longitudinal $\delta$-function
\begin{equation}
\nabla_i \chi(\br,t)=\frac{1}{4\pi}\int\rd^3 \br'\left(\nabla_i\nabla'_j\frac{1}{|\br-\br'|}\right)A_j^{\rm gc}(\br', t)
\end{equation}
where the primed derivative acts only on the Green's function and not on $A_j^{\rm gc}$. After integrating by parts, we identify
\begin{equation}
\chi(\br,t)=-\frac{1}{4\pi}\int\rd^3\br'\frac{1}{|\br-\br'|}\boldsymbol{\nabla}'\cdot\mathbf{A}^{\rm gc}(\br',t).\label{eqn:GeneratingFunction}
\end{equation}
The generating function $\chi(\br,t)$ can be obtained directly by using the explicit form of the field operator $\mathbf{A}^{\rm gc}$ from Eq.~(\ref{eqn:AOperator}) and evaluating the integrals in Eq.~(\ref{eqn:GeneratingFunction}). Alternatively, we take the divergence of Eq.~(\ref{eqn:ChangeA}) followed by a time derivative and find that the scalar potential in the true Coulomb gauge $\phi^c=\dot{\chi}$ satisfies the Poisson equation
\begin{equation}
-\nabla^2\dot{\chi}(\br,t)=\frac{\sigma(\brp,t)}{\epsilon_0}\;\delta(z)\label{eqn:PoissonEquationForChi},
\end{equation}
with the surface charge density
\begin{eqnarray}
\sigma(\brp,t)=-2i\int\rd^2\bkp|\bkp|\hspace{4cm}\nonumber\\
\times\left\{\left[\int_0^\infty \rd k_{zd}\sqrt{\frac{\hbar\epsilon_0}{2\w_\bk}}\;\hat{a}^{L}_{\bk\TM}(t)g_{\bk}^L(\brp)-{\rm H.C.}\right]\hspace{1 cm}\right.\nonumber\\
\left.+\left[\int_0^\infty\rd k_{z}\sqrt{\frac{\hbar\epsilon_0}{2\w_\bk}}\;\hat{a}^{R}_{\bk\TM}(t)g_{\bk}^R(\brp)-{\rm H.C.}\right]\right\}.\;\;\label{eqn:Sigma}
\end{eqnarray}
Here we have introduced the two mode functions
\begin{eqnarray}
g^R_\bk(\brp)&=&\frac{1}{(2\pi)^{3/2}}\frac{n^2-1}{2n^2}\left(1+R^R_\TM\right) e^{i\bkp\cdot\brp},\label{eqn:GL}\\
g^L_\bk(\brp)&=&\frac{1}{(2\pi)^{3/2}}\frac{n^2-1}{2n^2}
\frac{T^L_\TM}{n}e^{i\bkp\cdot\brp},\;\;\;\label{eqn:GR}
\end{eqnarray}
with reflection coefficients as given by Eqs.~(\ref{eqn:Frnl}). The solution of Eq.~(\ref{eqn:PoissonEquationForChi}) can be easily found as
\begin{eqnarray}
\dot{\chi}(\br,t)&=&i\int\rd^2\bkp e^{-|\bkp||z|}\nonumber\\
&\times&\left\{\left[\int_0^\infty \rd k_{zd}\sqrt{\frac{\hbar}{2\epsilon_0\w_\bk}}\;\hat{a}^{L}_{\bk\TM}(t)g_{\bk}^L(\brp)-{\rm H.C.}\right]\right.\nonumber\\
&\;&+\left.\left[\int_0^\infty\rd k_{z}\sqrt{\frac{\hbar}{2\epsilon_0\w_\bk}}\;\hat{a}^{R}_{\bk\TM}(t)g_{\bk}^R(\brp)-{\rm H.C.}\right]\right\}.\nonumber\\
\label{eqn:Phi}
\end{eqnarray}
As anticipated, the potential $\phi^c=\dot{\chi}$ turns out to be a second-quantized operator. It relates the vector potential in true Coulomb gauge to that in generalized Coulomb gauge via Eq.~(\ref{eqn:ChangeA}). It only affects photons with TM polarization and, interestingly, it is symmetric with respect to the interface i.e. $\dot{\chi}(-z)=\dot{\chi(z)}$. The generating function $\chi$ is found by integrating Eq.~(\ref{eqn:Phi}) with respect to time,
\begin{eqnarray}
\chi(\br,t)&=&-\int\rd^2\bkp e^{-|\bkp||z|}\nonumber\\
&\times&\left\{\left[\int_0^\infty \rd k_{zd}\sqrt{\frac{\hbar}{2\epsilon_0\w^3_\bk}}\;\hat{a}^{L}_{\bk\TM}(t)g_{\bk}^L(\brp)+{\rm H.C.}\right]\right.\nonumber\\
&\;&+\left.\left[\int_0^\infty\rd k_{z}\sqrt{\frac{\hbar}{2\epsilon_0\w^3_\bk}}\;\hat{a}^{R}_{\bk\TM}(t)g_{\bk}^R(\brp)+{\rm H.C.}\right]\right\}.\nonumber\\
\label{eqn:chi_result}
\end{eqnarray}

Let us now come back to the issue of the commutation relations between the field operators. In true Coulomb gauge we expect
\begin{eqnarray}
\left[A_i^{\rm c}(\br),\epsilon_0 E_j(\br')\right]&=&-i\hbar\delta^\perp_{ij}(\br-\br')=-i\hbar\delta_{ij}\delta^{(3)}(\br-\br')\nonumber\\
&+&i\hbar\nabla_i\nabla'_jG^0(\br-\br')\label{eqn:CommTrueGauge}
\end{eqnarray}
which is a consequence of the fact that $\boldsymbol{\nabla} \chi$ is the longitudinal part of $\mathbf{A}^{gc}$, cf. Eq. (\ref{eqn:ChiConstr}). This can also be confirmed by an explicit calculation using the mode functions (\ref{eqn:GL})--(\ref{eqn:GR}). The commutator splits as follows
\begin{eqnarray}
\left[A_i^{\rm c}(\br),\epsilon_0 E_j(\br')\right]=\left[A_i^{\rm gc}(\br)-\nabla_i\chi(\br),\epsilon_0 E_j(\br')\right]\hspace{11mm}\nonumber\\
=-i\hbar\delta^\epsilon_{ij}(\br,\br')-\left[\nabla_i\chi(\br),\epsilon_0 E_j(\br')\right]\hspace{3mm}
\label{eqn:SCGaugeComm}
\end{eqnarray}
where $\delta^\epsilon_{ij}(\br,\br')$ is given by Eq.~(\ref{eqn:GeneralizedDeltaTot}) and the reader is reminded that we consider the case $z'>0$ only. Substituting the mode functions (\ref{eqn:GL})--(\ref{eqn:GR}) into Eq.~(\ref{eqn:SCGaugeComm}), we find, using the same techniques as in the calculation of the completeness relation (\ref{eqn:Completeness}), that
\begin{eqnarray}
\left[\nabla_i\chi(\br),\epsilon_0 E_j(\br')\right]\hspace{55mm}\nonumber\\
=i\hbar\nabla_i\nabla'_j
\;\left\{ 
\begin{array}{lr}
-\dfrac{n^2-1}{n^2+1}\;G^0(\br-\br')& \mbox{ for }z<0,z'>0,\\
\\
\;\;G^R(\br,\br') & \mbox{ for }z>0, z'>0,
\end{array}
\right. \nonumber\\
\label{eqn:ChiECommutator}
\end{eqnarray}
where $G^0$ and $G^R$ are the Green's functions as introduced in Eqs.~(\ref{eqn:Poisson}) and (\ref{eqn:ReflectedPoisson}). Equation (\ref{eqn:ChiECommutator}) when combined with Eqs.~(\ref{eqn:GeneralizedDeltaTot}) and (\ref{eqn:SCGaugeComm}) confirms the assertion stated by Eq.~(\ref{eqn:CommTrueGauge}).

The above considerations have clearly demonstrated that the commutator between the vector potential and the electric field operators is gauge dependent. Therefore, the modification of the QED commutation relations is not a physical effect but rather is related to the choice of gauge in which the electromagnetic field is quantized, which is of course ultimately only a matter of convenience. However, we note that the commutation relations between the physical fields retain the standard form, as they should. Consider the commutator
\begin{equation}
\left[\mathbf{B}(\br),\mathbf{E}(\br')\right]=\curl\left[\mathbf{A}(\br),\mathbf{E}(\br')\right].\label{eqn:GaugeIndepComm}
\end{equation} 
We see from Eq.~(\ref{eqn:ChangeA}) that, regardless of the gauge one uses to calculate the right-hand side of the above relation, the end result is the same. The commutators (\ref{eqn:CommGenGauge}) and (\ref{eqn:CommTrueGauge}) differ only by a longitudinal part that is annihilated by the curl operator. Thus, the shape of the cavity has no impact on the fundamental commutation relations of physical fields.

\section{Perfect reflectors}
If the walls of the cavity are modelled as perfectly reflecting mirrors, the generalized Coulomb gauge (\ref{eqn:Generalized}) is meaningless. Then, a common way to quantize the electromagnetic field is to work with the free-space form of Eq. (\ref{eqn:Wave2}) in true Coulomb gauge (\ref{eqn:TrueCoulomb}) and demand that the fields are excluded from interior of the perfect reflector, i.e. one solves
\begin{eqnarray}
\left(\nabla^2-\frac{\partial^2}{\partial t^2}\right)\mathbf{A}(\br,t)&=&0,\nonumber\\
\diw\mathbf{A}(\br,t)&=&0,\label{eqn:PREquations}
\end{eqnarray}
together with the condition that the electric field vanishes for $z\leq 0$. This implies in particular that
\begin{equation}
E_x(z=0^+)=0,\ \ E_y(z=0^+)=0.
\end{equation}
The relation between the vector potential and the electric field is taken to be 
\begin{equation}
\mathbf{E}(\br,t)=-\frac{\partial \mathbf{A}(\br,t)}{\partial t},\label{eqn:AasE}
\end{equation}
and for this reason the boundary conditions for the electric field immediately imply rules for the vector potential. This method of quantization gives the vector field operator that can be be obtained by taking the $n\rightarrow\infty$ limit of Eq.~(\ref{eqn:AOperator}). This in turn implies that the commutation relations for the field operators are given by the perfect reflector limit of the commutation rule (\ref{eqn:CommGenGauge}) and \emph{not} by Eq. (\ref{eqn:CommTrueGauge}). Explicitly:
\begin{eqnarray}
\left[A_i(\br),\epsilon_0 E_j(\br')\right]=-i\hbar\delta_{ij}\delta^{(3)}(\br-\br')\hspace{2cm}\nonumber\\
+\frac{i\hbar}{4\pi}\nabla_i\nabla'_j \left(\frac{1}{|\br-\br'|}-\frac{1}{|\br-\bar{\br}'|}\right),\;\;\;z,z'>0,\;\;\label{eqn:CommGenGaugePR}
\end{eqnarray}
where $\bar{\br}=(x,y,-z)$. At first it seems surprising that, despite the Coulomb gauge condition having been imposed on the vector potential, the reflected part of the Green's function appears in the commutator. However, this can be explained as follows. In the presence of a perfect reflector the fluctuations of the quantized electromagnetic field imply the existence of a fluctuating charge density on the surface of the perfect reflector. Gauss's law reads
\begin{equation}
\diw\mathbf{E}(\br,t)=\frac{\sigma(\brp,t)}{\epsilon_0}\delta(z),\label{eqn:GaussLawPR}
\end{equation}
where $\sigma(\brp)$ is given as a perfect-reflector limit of Eq.~(\ref{eqn:Sigma}). Relation (\ref{eqn:GaussLawPR}) is a consequence of the boundary conditions applied to the electric field at $z=0$ (and vice versa). We observe that Eqs.~(\ref{eqn:PREquations}), (\ref{eqn:AasE}) and (\ref{eqn:GaussLawPR}) cannot be simultaneously satisfied on the surface of the perfect reflector. Thus, the gauge condition in Eq.~(\ref{eqn:PREquations}) must for a perfect reflector be amended to read
\begin{equation}
\diw\mathbf{A}(\br,t)=0\;\;\;\mbox{ for }\;z\neq 0
\end{equation}
which is in fact an adaptation of the generalized Coulomb gauge condition (\ref{eqn:Generalized}) to the case of the perfect reflector rather than the true Coulomb gauge. This is the origin of the reflected Green's function term appearing in the commutator (\ref{eqn:CommGenGaugePR}) as has also been pointed out in Ref.~\cite{Bimonte}. Our analysis also permits us to observe that the oversimplified model of perfectly reflecting cavity walls obscures the fact that the form of the commutation relation is actually determined by the choice of gauge. While it is claimed in Ref.~\cite{Bimonte} that the commutator between the physical fields (\ref{eqn:GaugeIndepComm}) is affected by the cavity walls if one assumes them to be perfectly reflecting, we have clearly shown this to be an erroneous conclusion.

\section{Hamiltonians}
Quantum electrodynamics in the presence of dielectrics is most conveniently formulated in the generalized Coulomb gauge. The minimal-coupling Hamiltonian of a charged particle that is placed near dielectric half-space  and coupled to the quantized electromagnetic field may be written as \cite{Dalton}
\begin{eqnarray}
H^{gc}&=&\dfrac{\left[\mathbf{p}-q\mathbf{A}^{gc}(\br_0)\right]^2}{2m}\nonumber\\
&+&\frac{1}{2}\int\rd^3\br\left\{\epsilon_0\epsilon(z)\left[\frac{\partial\mathbf{A}^{gc}(\br)}{\partial t}\right]^2+\frac{\mathbf{B}^2(\br)}{\mu_0}\right\}\nonumber\\
&+&\frac{1}{2}\int\rd^3\br\epsilon_0\epsilon(z)\boldsymbol{\nabla}\phi^{gc}(\br)\cdot\boldsymbol{\nabla}\phi^{gc}(\br),
\end{eqnarray}
where $\br_0$ is the position of the particle. In the following, it will prove most convenient to write the Hamiltonian $H^f$ of the electromagnetic field in the form
\begin{equation}
H^f=\sum_{\bk,\lambda} \hbar\w_\bk\left(a^\dagger_{\bk\lambda} a_{\bk\lambda} +\frac{1}{2}\right).
\end{equation}
The integral involving the scalar potential $\phi^{gc}$ is a $c$-number and it contains the infinite self-energy of the particle $\Xi$ as well as the $z_0$-dependent electrostatic interaction between the dielectric and the charge
\begin{equation}
\frac{1}{2}\int\rd^3\br\epsilon_0\epsilon(z)\boldsymbol{\nabla}\phi^{gc}(\br)\cdot\boldsymbol{\nabla}\phi^{gc}(\br)=\Xi + V^{es},
\end{equation}
with
\begin{equation}
V^{es}=-\frac{q^2}{4\pi\epsilon_0}\frac{n^2-1}{n^2+1}\frac{1}{4z_0}.\label{eqn:Ves}
\end{equation}
Equation (\ref{eqn:Ves}) can be seen as an interaction energy of a static charge with its image in the dielectric, multiplied by a factor of $1/2$ because the image is not independent but a consequence of the charge \cite{Jackson}. Dropping the irrelevant self-energy of the particle $\Xi$, one can write the Hamiltonian $H^{gc}$ as
\begin{equation}
H^{gc}=\dfrac{\left[\mathbf{p}-q\mathbf{A}^{gc}(\br_0)\right]^2}{2m}+H^f+V^{es}.\label{eqn:HamGC}
\end{equation}
Perhaps the most instructive way of obtaining the Hamiltonian in true Coulomb gauge $H^c$ is by using the unitary transformation
\begin{equation}
H^c=e^{iS/\hbar}H^{gc}e^{-iS/\hbar}+i\hbar\left(\frac{d}{dt}e^{iS/\hbar}\right)e^{-iS/\hbar},\label{eqn:Unitary}
\end{equation}
with the operator S is given by
\begin{equation}
S(\br_0,t)=-q\chi(\br_0,t).
\end{equation}
The generating function $\chi(\br,t)$ is given by Eq.~(\ref{eqn:chi_result}) and now taken at the position of the particle $\br_0$. In what follows we set operators to be time-independent (adopting the Schr\"{o}dinger picture) so that the term containing the time derivative in Eq.~(\ref{eqn:Unitary}) does not contribute. Then, using the same methods as in the proof of the completeness relation (\ref{eqn:Completeness}), it is not difficult to show that
\begin{equation}
e^{iS/\hbar}\left[\mathbf{p}-q\mathbf{A}^{gc}(\br_0)\right]e^{-iS/\hbar}=\left[\mathbf{p}-q\mathbf{A}^{c}(\br_0)\right],\nonumber
\end{equation}
as well as
\begin{eqnarray}
e^{iS/\hbar}H^f e^{-iS/\hbar}&=&H^f+\frac{i}{\hbar}\left[S(\br_0),H^f\right]\nonumber\\
&&+\frac{1}{2}\left(\frac{i}{\hbar}\right)^2\left[S(\br_0),\left[S(\br_0),H^f\right]\right]\nonumber\\
&=&H^f+q\dot{\chi}(\br_0)-\frac{n^2-1}{2n^2}V^{es}.\label{eqn:UnitaryDone}
\end{eqnarray}
With this, we obtain for the Hamiltonian in the Coulomb gauge
\begin{equation}
H^{c}=\frac{\left[\mathbf{p}-q\mathbf{A}^{c}(\br_0)\right]^2}{2m}+H^f+q\dot{\chi}(\br_0)+\left(\frac{n^2+1}{2n^2}\right)V^{es}.
\end{equation}
We see that compared to the Hamiltonian of Eq. (\ref{eqn:HamGC}) written out in the generalized Coulomb gauge, some of the electrostatic interaction energy has been redistributed and is now contained in the second-quantized part of the Hamiltonian $H^c$. One can actually see that this electrostatic interaction energy is now shared between two terms
\begin{equation}
H^{es}_{int}=q\dot{\chi}(\br_0)+\left(\frac{n^2+1}{2n^2}\right)V^{es}.\label{eqn:TwoTerms}
\end{equation}
Using standard time-independent perturbation theory applied to the interaction term $q\dot{\chi}(\br)$, one finds that the first non-vanishing contribution is of second order in the perturbation and is given by
\begin{eqnarray}
\Delta E^{es}&\!\!=&\!\!\sum_{\bk,\mathbf{p}_f}\dfrac{\left|\langle \mathbf{p}_f ;1_{\bk\TM}|q\dot{\chi}(\br_0)|\mathbf{p};0\rangle\right|^2}{\dfrac{\mathbf{p}^2}{2m}-\left(\dfrac{\mathbf{p}_f^2}{2m}+\w_\bk\right)}
\approx -q^2\sum_\bk\dfrac{\left|\dot{\chi}(\br)\right|^2}{\w_\bk}\nonumber\\
&\!\!=&\!\!-\frac{q^2}{2\epsilon_0}\int\rd^2\bkp e^{-2|\bkp | z_0}\nonumber\\
&&\times\left[\int_0^\infty\rd k_{zd}\dfrac{\left|g^L_\bk(\brp)\right|^2}{\w_\bk^2}+\int_0^\infty\rd k_{z}\dfrac{\left|g^R_\bk(\brp)\right|^2}{\w_\bk^2}\right],\nonumber\\ \label{eqn:SecondOrderShift}
\end{eqnarray}
where we have used the no-recoil approximation. The mode functions $g$ are given in Eqs.~(\ref{eqn:GL})--(\ref{eqn:GR}). The resulting integrals in Eq.~(\ref{eqn:SecondOrderShift}) can be calculated analytically and the result is
\begin{equation}
\Delta E^{es}=\left(\frac{n^2-1}{2n^2}\right)V^{es}.
\end{equation}
Thus, the contributions from both terms in Eq.~(\ref{eqn:TwoTerms}) add up to yield the whole of the electrostatic interaction energy
\begin{equation}
\left(\frac{n^2-1}{2n^2}\right)V^{es}+\left(\frac{n^2+1}{2n^2}\right)V^{es}=V^{es}. 
\end{equation}
This is of course what one would expect since both formulations of the theory must lead to the same physical results.

\section{Conclusions}
In this paper we have illustrated some intricacies involved in the quantization of the electromagnetic field when polarizable boundaries are present and modelled macroscopically by the introduction of the spatially varying and piecewise constant dielectric function. Starting from the generalized Coulomb gauge we have derived the expression for the coordinate representation of the unit kernel in that gauge, thereby explicitly verifying the completeness relation of the mode functions. While this calculation has its own merit, it has served us to develop tools that allow us to explicitly carry out a gauge transformation from the generalized Coulomb gauge to the true Coulomb gauge, where the expression for the vector field operators is truly transverse even in the presence of the boundaries. This has shed light on some misconceptions about the nature of the commutation relations in macroscopic quantum electrodynamics, especially in the case when the boundaries are modelled as perfect reflectors. 

We have also written down the Hamiltonian for a charged particle near a dielectric boundary in true Coulomb gauge and shown that and why it is different from the one in generalized Coulomb gauge. It contains extra terms due to an induced fluctuating surface charge at the boundary, now represented as a second-quantized operator. This term contains parts of the electrostatic interaction of a particle and the surface, which in generalized Coulomb gauge is represented by a c-number, namely the electrostatic potential obtained by classical methods, e.g. the method of images. Finally, we have explicitly demonstrated the gauge invariance of the theory by working out the electrostatic parts of the charge-surface interactions. This work paves the way to more elaborate gauge transformations which provide a link between well understood approaches to macroscopic QED and more elaborate theories.


\begin{thebibliography}{99}
\bibitem{maximumPrincipleFootnote}{We aim here to make the left-hand side of Eq.~(\ref{eqn:ChangeA}) transverse. Setting $\boldsymbol{\nabla}\chi=\mathbf{A}^{gc}_\parallel$ does the job, but so would
\begin{equation}
\boldsymbol{\nabla}\chi=\mathbf{A}^{gc}_\parallel + \mathbf{f}\label{eqn:f}
\end{equation}
where $\mathbf{f}$ is any function with vanishing divergence i.e. $\diw \mathbf{f}=0$. To convince ourselves that $\mathbf{f}$ necessarily vanishes we note that Eq.~(\ref{eqn:f}) implies that $\boldsymbol{\nabla}\times \mathbf{f}=0$. This in turn implies that $\mathbf{f}$ must be the gradient of some harmonic function $\Phi$ i.e. satisfying $\nabla^2\Phi=0$. Since we expect the fields to vanish at infinity the function $\Phi$ is necessarily zero by the maximum principle, see e.g. \cite{Gamelin}.}
\end{thebibliography}
\end{document}